# Relativity Theory and a Real Pioneer Effect

*Peter Ostermann**

Keeping the relativistic laws of motion a non-conventional Pioneer effect would prove an increase of the scale rate of atomic clocks in comparison with planetary ones. Together with a slowly decreasing amount of about 60% due to anisotropic radiation this would be a thinkable explanation for an apparent anomalous acceleration of the Pioneer 10/11 probes. Such a difference between atomic time and ephemeris time, however, (coincidentally corresponding to that of atomic time and cosmic time as derived from a cosmological model of general relativity, where the cosmic 'coordinate' speed of light is fixed to c* = c) is ruled out by solar system's observational facts. Thus a non-conventional Pioneer effect would inevitably contradict relativity theory.

*A) Situation* – As commonly known, Anderson *et al.* [1] reported an apparent constant anomalous acceleration $a_P$ = $(8.74 \pm 1.33) \times 10^{-8}$ cm/s$^2$ of the Pioneer 10/11 spacecraft towards the Sun, which is not observed for Moon and planets. Since it appears that Pioneer 10 has sent its last signal to Earth, it may be the time to pay attention once more to this Pioneer effect. At least a considerable amount of the observed effect should have its origin in the spacecraft's anisotropic radiation considered first by Katz [2] and Murphy [3]. Although Scheffer's [4] corresponding model looks most likely indeed, there might remain some constant 'anomalous acceleration' indicated by the model's prediction of *"a larger decrease in acceleration ... than is observed"*. On the other hand, Anderson and his co-authors merely included those anisotropic radiation effects into the relative error range of about ± 15%.

In a double equation [1]/(15) of their last paper [written here as (1), (2), avoiding an unusual sign convention] the Pioneer authors gave the basic result of their measurements carried out and evaluated with utmost care

$$f_{obs} - f_{model} \approx \frac{2 a_P t}{c} f_{reference}, \qquad (1)$$

where $f$ are frequencies, $t$ is the ephemeris time and $c$ the speed of light. In the Doppler data $f_{obs}$ of the two-way return signals observed by atomic clocks this actually looks like a blue shift with respect to the modeled frequencies

$$f_{model} \approx f_{reference}\left(1 - \frac{2 v_{model}(t)}{c}\right). \qquad (2)$$

Here $v_{model}(t)$ is calculated using the relativistic laws of motion. Over a reasonable span of the Pioneer-10 data it is even sufficient to do these calculations approximately on Newton's laws taking into account the solar radiation pressure, as Markwardt [5] carried out in an independent analysis.

However, the influence of the anisotropic thermal radiation stated above should lead to a correction from $a_P$ to

$$a_{P-0} = (1 - \Theta) a_P. \qquad (3)$$

Here the parameter $0 < \Theta < 1$ may represent an average value for that slowly decreasing fraction $\Theta(t)$ which corresponds to the thermal recoil acceleration. The energy supply of the probes was provided by $^{238}$Pu with a half-life of 88 years. So even the evaluated period of observation 1987–1998 for Pioneer 10 extends over only 1/8 of that value, which makes it difficult to separate the thermal contribution $a_P \Theta(t)$ from a non-conventional approximately constant amount $a_{P-0}$.

*B) Preliminaries* – From a purely logical view it is necessary to distinguish atomic time (though in various forms) from at least two other *natural* time scales. It is possible to define them without reference to each other:

a) Independent of any theory of gravitation, understand the *ephemeris time t* to be that scale on which the motion of a single ideal planet would show constant periods from periastron to periastron in the field of an ideal star (except for negligible gravitational radiation). In the following the coordinate time of the solar system is fixed uniquely to be this ephemeris time $t$, indicated by 'planetary clocks'.

b) Independent of any theory of gravitation, understand the *cosmic time* $t^*$ to be that scale on which far away from local gravitational sources the cosmic coordinate speed of light is constant $c^* = c$. In the following the coordinate time of the universe is fixed uniquely to be that cosmic time $t^*$, established by intergalactic signals of light.

In this paper on hand a real Pioneer effect $a_{P\text{-}0}$ is considered to be constant (within all achievable accuracy) and compatible to the 'geodesic' equations of motion $\delta \int ds = 0$ of general relativity theory (GRT). As usual the local line element $ds$ is determined by the local energy-stress tensor, i.e. by Einstein's equations $E_{ik} \equiv R_{ik} - \tfrac{1}{2} R g_{ik} = \kappa T_{ik}$ ($i, k = 0..3$) implying the law of inertia far away from local sources. The equivalence principle holds as follows: *In a freely falling local frame (even including self-gravitating bodies) no influence is measurable of the gravitational potential, which may cause an acceleration of the frame as a whole.*

*C) Exploration* – Now one may ask the following questions: What would have been measured after a short braking maneuver at a large distance from the Sun, which had slowed down the speed of the probes to zero? What in case of a return to Earth? What finally, if the probes had not started at all?

If a real constant Pioneer effect is strictly thought to exist, then relations (1), (2) must hold for these situations, too. – So let us consider

$$v_{\text{model}}(t) = 0. \tag{4}$$

Now (2) yields

$$f_{\text{model}} = f_{\text{reference}}, \qquad (v_{\text{model}} = 0) \tag{5}$$

and with the correction (3) from $a_P$ to $a_{P\text{-}0}$ one gets from relation (1)

$$f_{\text{obs}} \approx f_{\text{reference}} \left(1 + 2 \frac{a_{P\text{-}0}}{c} t\right). \qquad (v_{\text{model}} = 0) \tag{6}$$

To date, in view of an observer looking from the solar system barycenter (SSB) the frequency $f_{\text{reference}}$ should obviously be that of a terrestrial transmitter controlled by an atomic clock (maser) and follow from the conventional understanding of relativity theory. Thus, taking into account the influences of local gravitational potential and velocity, using a frequency $f^*$ which is constant with respect to ephemeris time $t$, one has

$$f_{\text{reference}} = f^* \frac{ds}{c\,dt} \approx f^* \left(1 + \frac{U}{c^2} - \frac{1}{2}\frac{v^2}{c^2}\right), \tag{7}$$

where $U$ is the (negative) Newtonian potential in which a terrestrial atomic clock may be stationed, and $v$ the velocity of the clock relative to SSB.

Understand $\sigma$ (usual name 'proper time') to be a variable for the display of such an atomic clock, with $\Delta\sigma$ the number $\Delta Z_\sigma$ of ticks $T_\sigma$ (multiplied by an arbitrary constant unit $T^*$ of time). Now, consider a maser, controlling a transmitter whose frequency $f_\sigma$ increases according to (6): *It is clear that the same maser, if used as an atomic clock, will display increasing numbers $\Delta Z_\sigma$ of (decreasing) periods $T_\sigma(t) = 1/f_\sigma(t)$, whenever measuring constant intervals $\Delta t$ of ephemeris time in the future.* This means the scale rate of an atomic clock

$$\frac{d\sigma}{dt} = \frac{f_\sigma}{f^*} \tag{8}$$

would increase with ephemeris time $t$. Thus, with $f_\sigma \equiv f_{\text{obs}}(v_{\text{model}} = 0)$ using (6), (7) one gets

$$d\sigma \approx dt \left(1 + 2\frac{a_{P\text{-}0}}{c} t\right)\left(1 + \frac{U}{c^2} - \frac{1}{2}\frac{v^2}{c^2}\right). \tag{9}$$

This differs by the factor $(1+2a_{P-0}t/c)$ from relation [1]/(4) of the Pioneer paper (where TAI ≡ $\sigma$, ET ≡ $t$, and $U$ signed positive). Therefore, the integration of relation [1]/(4) leading to their formula [1]/(5) between Ephemeris Time $ET$ and International Atomic Time $TAI$ has to be corrected approximately by $-(a_{P-0}/c)t^2$. This gives

$$ET - TAI \approx \Delta_{conv} - \frac{a_{P-0}}{c}(ET)^2, \qquad (10)$$

where for the sake of shortness all conventional terms (constant or periodical) of the original relation are summarized as $\Delta_{conv}$. Decisive, however, is the additional monotonously variable term on the right: though both being natural, barycentric atomic clocks and planetary clocks would show different times.

*D) Coincidence* – Preliminary b) demands that *every* spatial Euclidean cosmic line element is to be written in a simple scalar form as follows:

$$d\bar{s}^2 = \zeta^2 \left\{ c^2 dt^{*2} - dl^{*2} \right\}, \qquad (11)$$

where $\zeta \approx 1 + Ht^* + O^2(Ht^*)$, with $H$ Hubble's constant, and a bar here on $d\bar{s}$ ($\equiv c\,d\bar{\sigma}$) indicates the spatial Euclidean large scale average (as opposed to $ds$ for the local element). This general scalar form is fixed uniquely by the requirement of a constant intergalactic speed of light $c^* = c$ [note that $c^* = c$ is *not* given in any form other than (11)]. Now, with the special assignment

$$\zeta = e^{Ht^*}, \qquad (12)$$

the cosmic scalar form (11) together with its Einstein tensor $\bar{E}_{ik}$ not only turns out to be non-singular, but strictly *stationary* as well (not static of course): because of the exponential form of the time scalar $e^{Ht^*}$, all the resulting relative temporal changes always depend solely on *differences* $\Delta t^* = t^* - t_0^*$. This allows to set any reference point of time to be $t_0^* = 0$ for arbitrary complexes of observation. Therefore, choosing appropriate units respectively, no special point (but a direction) of the cosmic time scale is preferred.

According to the basics of general relativity, the display of an intergalactic atomic clock is understood to be $d\bar{\sigma} = d\bar{s}/c$. Thus a redshift $z = e^{Hd/c} - 1$ of intergalactic electromagnetic waves (emitted from sources at rest with respect to the cosmic microwave background) is immediately concluded from (11) with $d = ct$ the covered distance. This must occur since on the one hand, due to the intergalactic speed of light $c^* = c$, the oscillation period of the starlight remains constant during its propagation with respect to ephemeris time $t$. On the other hand, the same oscillation period must have increased (redshifted) with respect to the decreasing periods $T_\sigma(t) = T^* e^{-Ht^*}$ of intergalactic atomic clocks, because these decreasing periods actually represent those of 'new' photons at time and place of their origin.

With regard to (11), (12), Einstein's equations may be written in the form

$$\bar{E}_{ik} = \begin{pmatrix} \frac{2H^2}{c^2} & 0 & 0 & 0 \\ 0 & 0 & 0 & 0 \\ 0 & 0 & 0 & 0 \\ 0 & 0 & 0 & 0 \end{pmatrix} - \bar{p}\,\bar{g}_{ik} = \kappa \bar{T}_{ik}, \qquad (13)$$

demanding a negative gravitational pressure $\bar{p} = -1/3 \rho_c e^{-2Ht^*}$ where $\rho_c = 3H^2/(\kappa c^2)$ is the critical density. Obviously $\bar{p}$ corresponds to a stationary changing 'cosmological constant'. To state here explicitly, the gravitational pressure *must* be negative because the walls of a large-scale box including a plenty of galaxies statistically at rest, would have to pull *outwards*, if those inside should not mass together after those outside would have been removed. With respect to (13) the phenomenological mass density should be only $2H^2/c^4 = (2/3)\rho_c/c^2$.

From (11) one has $d\bar{\sigma} = \zeta\, ds_{SRT}/c$. Now, if the coordinate time $t$ of the solar system (ephemeris time) was the *same* as the coordinate time $t^*$ of the universe (cosmic time) then it would be consequent to conclude $d\sigma = \zeta\, ds_{GRT}/c$ for the display of an arbitrarily moving atomic clock in a local gravitational potential, which is described by the line element $ds \equiv ds_{GRT}$. Though in contrast to the commonly accepted understanding, that the line element $ds$ should be identical with the display of an atomic clock $d\sigma$, this would directly follow from the equivalence principle (a *cosmic* violation of the identity is not unthinkable to date). With the stationary time scalar $\zeta = e^{Ht^*}$ one then concludes

$$d\sigma = e^{Ht^*}\frac{ds}{c} . \qquad (14)$$

Analogous to (8) the cosmic influence on the scale rate of atomic clocks has to be included in their frequency, too, multiplying the time factor $\zeta$ to the well-known influences of local gravitational potential and velocity. Therefore, (using $t^* \equiv t$) from (14) together with (7), (8) one gets

$$f_\sigma \approx f^* \frac{d\sigma}{dt} \approx e^{Ht} f_{\text{reference}}. \qquad (15)$$

Comparing (6) and (15) one would find $2a_{P-0}/c$ to be Hubble's constant $H$ indeed:

$$H \stackrel{!}{\approx} \frac{2a_{P-0}}{c} = (1-\Theta)\frac{2a_P}{c} \approx (1-\Theta) \times 180\,\frac{\text{km/s}}{\text{Mpc}}. \qquad (16)$$

With $H = 2a_{P-0}/c \approx 70$ km/s/Mpc according to [6] the average thermal part of the measured value $a_P$ should be about $\Theta \approx 60\%$. Now relation (9), concluded from a real Pioneer effect, corresponds to (14) with $t^* := t$, thus explaining the anomalous Doppler shift (6). Remember that this relation would result from the cosmic scalar form (11), if the ephemeris time $t$ of the solar system was the same as the cosmic time $t^*$ of the universe. Therefore, inserting $U = 0$, $v = 0$ into (9), a redshift $z \approx Hd/c$ would directly follow from a real Pioneer effect.

For a given law of motion the tracking of spacecraft in a known gravitational field definitely establishes a *natural* ephemeris time $t$ (*"The spacecraft acceleration is integrated numerically to produce the spacecraft ephemeris."* [1]). Thus, if there was a real effect, the Pioneers in their changing positions would have been used as *clock hands of the cosmic time $t^* = t$*.

*E) Discussion* – The simplest explanation for the redshift of starlight is to make a difference between atomic time $\sigma$ and cosmic time $t^*$. But the cosmic effect on atomic clocks considered here (which could be the reason for spontaneous emission or decay processes, too) would only explain a real Pioneer effect, if again ephemeris time $t$ was the same as cosmic time $t^*$. – But it is *not*.

A substitution [1]/(61) of the form $t \to t + (a/2)\,t^2$ corresponding to (10), or alternatively a frequency drift [1]/(62) corresponding to (6) were already considered by the Pioneer authors as well as other phenomenological time models. In spite of good Doppler fits, they were rightly rejected at last.

According to (10), the increasing quadratic difference of barycentric atomic time to ephemeris time arisen since the definition of the SI-second in 1969 would amount to little more than about 1.3 s. This, however, would mean displacements of about 39 km, 32 km and 2.6 km in the positions of the Earth, Mars and the Moon respectively, where up to 140 m would get into the Moon's radial component because of the lunar orbit's eccentricity. Thus, such displacements are huge in comparison to corresponding observational accuracies, i.e. a decameter to few meter Mars ranging accuracy and (at present) a centimeter lunar ranging accuracy, where the observation times of the Moon and the planets are recorded with atomic clocks. Furthermore, using atomic time and length e.g. in Kepler's third law would result in an effectively varying gravitational constant $G_{\text{eff}} = G_0\,e^{-Ht}$, if there was a corresponding difference between atomic time and ephemeris time. This is ruled out by decisive *G*-dot results like [7],[8], where the Hellings *et al.* paper obviously already considered and tested the idea that atomic clocks and gravitational clocks might be deviating from one another. Therefore, a deviation of barycentric atomic time from ephemeris time as discussed above is to be excluded [9] and thus a non-conventional Pioneer effect would inevitably contradict relativity theory. In this view, the observed effect should have its origin entirely in conventional physics like anisotropic radiation (as already indicated above, see A) and the reported value of $a_P/c \approx H$ seems to be nothing but a strange numerical coincidence.

*F) Appendix* – A first corresponding attempt on the Pioneer effect was given in a previous paper [10] treating the relativistic cosmic scalar form (11) in much more detail and showing a stationary universe to be in harmony with important observational facts. In particular, with the redshift parameter $z = e^{Hl/c} - 1$ for galaxies statistically at rest, this led to a relation for the apparent luminosity of galaxies $\propto [(1+z) \ln(1+z)]^{-2}$ obviously near the impressive magnitude versus redshift observations reported in [6] (these observations are interpreted today as a cosmic acceleration, thereby demanding *ad-hoc* constructs like 'quintessence' or 'dark energy'). In spite of an infinite number of galaxies, another result is a finite average stellar radiation density equivalent in its integral to that of a 3-6 K black body radiation, as well as a maximum in the numerical distribution of galaxies versus *z*. – Alternatively to the explanation of a real Pioneer effect given there, it was shown that using (14) as an cosmic *embedded* line element (with the equations of motion now $\delta \int d\sigma = 0$ instead of $\delta \int ds = 0$), the scale rate of ephemeris time *t* would be approximately equal to barycentric atomic time $\sigma_{SSB}$ and hence no real Pioneer effect would result from that.

\* Electronic address: <box@peter-ostermann.de>.

___________________________________